\documentstyle[prl,aps,epsf]{revtex} 
\draft 
\begin{document}
\twocolumn[\hsize\textwidth\columnwidth\hsize\csname @twocolumnfalse\endcsname

\title{Phase-Locking of Vortex Lattices Interacting with Periodic Pinning} 
\author{Charles Reichhardt, Richard T. Scalettar, and  Gergely T.\ Zim\'{a}nyi}
\address{Department of Physics, University of California, Davis, California 95616.}

\author{Niels Gr{\o}nbech-Jensen}
\address{Department of Applied Science, University of California, 
Davis, California 95616.\\
NERSC, Lawrence Berkeley National Laboratory, Berkeley, California 94720.}

\date{\today}
\maketitle
\begin{abstract} 
We examine Shapiro steps for  
vortex lattices interacting with periodic pinning arrays 
driven by AC and DC currents.
The vortex flow occurs by the motion of the interstitial vortices
through the periodic potential generated by the vortices that remain
pinned at the pinning sites. Shapiro steps are observed for 
fields $B_{\phi} < B < 2.25B_{\phi}$ with the most pronounced steps 
occurring for fields where the interstitial vortex lattice has a high 
degree of symmetry.     
The widths of the phase-locked current steps as a function of the
magnitude of the AC driving are found to follow a Bessel function in
agreement with theory.
\end{abstract}

\vskip2pc]
\narrowtext
 
Vortex lattices interacting with periodic pinning arrays show a wide
range of interesting commensurability or matching effects when the number of 
vortices is a multiple or rational-multiple of the number of pinning sites.   
These pinning arrays can be created with lithographic techniques in which  
arrays of microholes or "antidots" \cite{Metlushko,Moschalkov,Harada} 
and magnetic dot arrays \cite{Schuller} can act as pinning sites. 
For small pinning sites only one vortex is trapped on a site
as observed in transport measurements \cite{Moschalkov}, 
Lorentz-microscopy experiments \cite{Harada} and simulations
\cite{Reichhardt}. Additional vortices sit in the areas between the pins  
and under the influence of an applied driving force they can 
flow between the vortices that have remained trapped at 
the pinning sites \cite{Moschalkov,Harada,DrivenShort} 
The flowing interstitial vortices experience a periodic potential 
caused by the repulsive interactions from the vortices at the
pinning sites. The motion of the driven interstitial vortices is
then analogous to an over-damped particle moving down a tilted washboard.
With the addition of an AC driving current, interference effects in 
the form of Shapiro steps can be expected to occur when the 
frequency of the particles moving over the washboard matches with one of the
harmonics of the driving frequency \cite{Shapiro}.
Recently Shapiro steps have been observed for 
driven vortices moving in samples with a periodic array of 
pinning sites at twice the 
matching field $B = 2B_{\phi}$ \cite{Bael} where $B_{\phi}$ is the field
for which there is one vortex per pinning site. 
The height of these current steps (range of phase-locking)
strongly suggests that the vortex motion consists of the interstitial vortices
moving in the periodic potential from the pinned vortices.
Shapiro steps have also been observed by Martinoli {\it et al.}
\cite{Daldini} for vortices moving over a one dimensional 
periodic potential created from a periodic thickness modulation. 
It has further been proposed that Shapiro steps can be seen for  
vortices in driven flux-transformers. 

In this work we investigate numerically and analytically the 
Shapiro steps for driven vortices in thin
film superconductors with periodic pinning arrays.
The vortex lattice consists of the
pinned vortices at the pinning sites and the sublattice of vortices that 
sit in the interstitial region. As a function of increasing drive we
observe the
interstitial vortices moving in one dimensional channels between the pinning sites. With 
a superimposed AC drive we observe Shapiro steps. We find that 
for certain commensurate fields, such as $B= 2B_{\phi}$, 
the system can be modeled as  
an overdamped driven pendulum with the associated phase locking.
We find numerically that the 
widths of the steps depend on the magnitude of the AC driving as a Bessel 
function in agreement with theory. The Shapiro steps are most 
pronounced for highly symmetric interstitial vortex lattice arrangements. For
$ B > 2B_{\phi}$ the steps vanish out due to complicated vortex
configurations leading to nontrivial flow patterns.
  
We numerically integrate the overdamped equation of motion for
a vortex $i$ 
\begin{equation}
{\bf f}_{i} = {\bf f}_{i}^{vv} + {\bf f}_{i}^{vp} + {\bf f}_{d} 
+ {\bf f}_{ac} = {\bf v}_{i}   \; .
\end{equation} 
Here the total force acting on vortex $i$ is ${\bf f}_{i}$.
The vortex-vortex interaction potential is 
logarithmic, $U_{v} = -\ln (r)$,
and the force on vortex $i$ from all the other vortices is
${\bf f}_{i}^{vv} = 
\sum_{j\neq i}^{N_{v}}\nabla_i U_{v}(r_{ij})$
\cite{Jensen1}. 
We impose periodic boundary conditions and evaluate the periodic long-range
logarithmic interaction with an exact and fast converging sum 
\cite{Jensen2}. 
The pinning is modeled as attractive parabolic wells with $f_{i}^{vp} = 
(f_{p}/r_{p})\Theta(|{\bf r}_{i} 
- {\bf r}_{k}^{(p)}|/\lambda){\hat {\bf r}}_{ik}^{(p)}$. Here, $\lambda=1$,
$\Theta$ is the step function, ${\bf r}_{k}^{(p)}$ is the location of
pinning site $k$, $f_{p}$ is the maximum pinning force and
${\hat {\bf r}}_{ik}^{(p)} = ({\bf r}_{i} - 
{\bf r}_{k}^{(p)})/| {\bf r}_{i} - {\bf r}_{k}^{(p)}|$. The pinning is placed
in a rectangular array ($L_x,L_y$)
with the ratio of the pinning radius $r_{p}$ to pinning 
lattice constant $L_y$ being $r_{p}/L_y =  0.164$, 
close to the ratio $0.2$ used in the experiments \cite{Bael}.
The pinning is placed in a $4\times4$ array 
and the initial vortex configurations are obtained by annealing from
a high temperature state where the vortices are liquid and cooling to
$T=0$. For certain parameters we 
have also considered simulations for
pinning arrays up to $ 10\times10$ and found only minor differences.  
We only consider the case for $B > B_{\phi}$ so that the vortex motion
will be strictly from the flow of interstitial vortices.  
The driving force ${\bf f}_{d}$ represents the Lorentz force from an
applied current. We gradually increment ${\bf f}_{d}$
from zero simulating each DC current value for
17500 time steps (the normalized time step is $dt=0.003072$)
to obtain the average 
of the vortex velocities. The resulting DC force-velocity curve is 
proportional to the DC current-voltage curve.
The AC offset is added as $f_a \cos(\omega t)$. We conduct a 
series of simulations where the amplitude $f_a$ is varied. In this 
work both the DC and AC driving forces will be in the $x$-direction. 

We first consider the $B = 2B_{\phi}$ case where the interstitial 
vortices form a perfectly ordered square sub-lattice. 
The vortex trajectories above depinning are shown in
the upper inset of Fig.\ 1 for this case.  Here the  
interstitial vortices travel in one dimensional paths between the 
pinned vortex sub-lattice. Further, 
the moving interstitial vortex lattice retains the same square 
symmetry as the pinned interstitial vortex lattice.  Fig.\ 1 shows 
typical simulation results of the voltage response
$V_{x} = (1/N_{v})\sum_{i=1}^{N_{v}}{\hat {\bf v}}_{i}\cdot{{\hat {\bf x}}}$  
versus an applied DC driving force at 
several different AC amplitudes for $B = 2B_{\phi}$. 
The simulation parameters are $\omega=1.6276$ and $L_x=L_y=1.83$. 
For zero AC driving the vortex velocities 
increase {\it linearly} with the DC driving force. With applied
AC driving there are clear steps 
where the vortex velocities remain constant for a finite range of 
DC driving, indicative of {\it phase-locking} of the vortex motion. 
The widths of the steps depend on the magnitude of the AC drive. 

In order to demonstrate that the phase-locking of the
interstitial vortex motion is
indeed closely related to the well-known Shapiro steps in the AC driven
pendulum equation \cite{Barone} we first make the  observation
from the inset in  Fig.\ 1 that the interstitial vortices are moving 
{\it one-dimensionally} along the x-direction at the symmetry line
between the pinned vortices $y=\frac{L_y}{2}$, where
$L_y$ is the distance between two pinning centers along the y-direction. 
This allows us to write the equation of motion for the unpinned vortices as,
\begin{eqnarray}
\frac{d}{dt}x_i - f_i^{vv}(x,y=\frac{L_y}{2}) & = & f_d + f_{ac} \; ,
\end{eqnarray}
where we have neglected the pinning interaction and motion in the transverse
direction. We will make the additional assumptions that unpinned
vortices form a perfect rectangular lattice, meaning that they effectively
do not interact due to symmetry, and that the pinned vortices are effectively
pinned exactly to their pinning site; i.e., that the pinned vortices have
no dynamics and form a perfect rectangular lattice with dimensions $L_x$ and
$L_y$.

Under these assumptions each moving vortex obeys the following
equation of motion \cite{Jensen2}:
\begin{eqnarray}
\frac{d}{dt}x_i - \frac{\pi}{L_x}\sum_{k}\frac{\sin(2\pi\frac{x}{L_x})}
{\cosh\left(2\pi\frac{L_y}{L_x}(k+\frac{1}{2})\right)-\cos(2\pi\frac{x}{L_x})}
& = & f_d + f_{ac} \; .
\end{eqnarray}
Considering only the leading term in the above sum, we can simplify the
interaction between pinned and unpinned vortices to yield the equation,
\begin{eqnarray}
\frac{d}{dt}x - \frac{2\pi}{L_x}{\rm sech}\left(\pi\frac{L_y}{L_x}\right)\sin(2\pi\frac{x}{L_x}) = f_d + f_a\cos(\omega t) \; ,
\end{eqnarray}
where we have considered only the contributions, $k=-1,0$, and allowed for
a relative error in the force of
$\sim{\rm sech}\left(\pi\frac{L_y}{L_x}\right)$, which is obviously
small as long as $L_y/L_x$ is not small. This equation describes the
driven overdamped pendulum, and we can therefore apply the procedure for
evaluating phase-locking ranges between a pendulum and an AC drive.

Assuming phase-locking where the pendulum (vortex) moves with a frequency
$n\omega$, we insert the following ansatz (valid for large AC amplitudes)
into the above equation,
$x(t) = x_0 + n\omega\frac{L_x}{2\pi}t + \frac{f_a}{\omega}\sin{\omega t} \; $,
and equate the DC components of the resulting expression, yielding the
relationship between the applied AC force and the phase, $2\pi{x_0}/L_x$,
for a given integer $n$:
\begin{eqnarray}
nL_x\frac{\omega}{2\pi} - \frac{2\pi}{L_x}{\rm sech}\left(\pi\frac{L_y}{L_x}\right)J_n\left(\frac{2\pi f_a}{\omega L_x}\right)\sin(2\pi\frac{x_0}{L_x}) & = & f_{d} \; ,
\end{eqnarray}
where $J_n$ is the $n$th order Bessel function of the first kind.
The size of the range, $\Delta f_d$, in $f_{d}$ for which the vortex 
motion
may stay locked to the AC drive's $n$th harmonic is then given by the extreme
values of $\sin(2\pi\frac{x_0}{L_x})$:
\begin{eqnarray}
\Delta f_{d} & = & \frac{4\pi}{L_x}{\rm sech}\left(\pi\frac{L_y}{L_x}\right)
\Big|J_n\left(\frac{2\pi f_a}{\omega L_x}\right)\Big| \; .
\end{eqnarray}

By conducting a series of simulations    
with different AC driving amplitudes we can compare our simulation results 
for the dependence of the step widths with those predicted from equation (6).
In Fig.\ 2a we plot the widths of the locking ranges for the
harmonics, $n = 0,1,2$, predicted for our parameters from equation (6)
(solid lines) and the widths of the simulated locking ranges, $n = 0$
($\bullet$), $n = 1$ ($\circ$), and $n=2$ ($\Box$).  
There is very good agreement between the simulation data and the 
predicted curves. We note that although equation (6) is for a single 
interstitial moving vortex at $B = 2B_{\phi}$ the interstitial vortex
lattice is symmetric (see Fig.\ 1a) so the interstitial-interstitial 
vortex interactions cancel. 
We also obtain good agreement for the predicted widths from equation (6) and 
the simulations for the higher harmonics $n > 2$  
which are not shown here. We
note that the agreement between the simulation data and the predicted
behavior is not expected to be exact since the force that the interstitial 
vortices experience from the pinned vortex lattice 
is not strictly sinusoidal. We have also tested equation (6) 
for different ratios of $L_{x}/L_{y}$ by 
considering a rectangular pinning array with $L_{x}/L_{y} = 2$.
The ratio 
of the step widths for the different directions is $\approx 52$ in 
good agreement with the theoretical prediction of $\approx 57$. 
The agreement is still good when we compare the simulated ranges
of phase-locking with those predicted by equation (6) for the same
parameters as above, but with $L_x=2L_y=3.66$. Since the vortices
are forced in the x-direction, this is a case where the harmonic
potential approximation made in equation (4) is not expected to
be as good as for the square lattice case, $L_x=L_y$. Figure
2b shows that simulations at the second matching field, $B = 2B_{\phi}$,
($\Box$ and $\circ$) show less than predicted ranges of locking suggesting
that assumptions in the analysis are not well within validity. However,
performing the same simulations, but at the matching field with {\it one}
additional interstitial vortex (filled markers), reveals locking-ranges
very close to what is predicted. This underlines that the harmonic
potential assumption made in equation (4) is reasonable even for
$L_x=2L_y$. Closer examination of the dynamics at $B = 2B_{\phi}$
(open markers) shows that internal modes in the moving vortex lattice
are being excited and the assumption of cancelation of interstitial
vortex interactions become invalid, which is responsible for the
deviation between simulations and our prediction in figure 2b for
$B = 2B_{\phi}$. It is, of course, important to emphasize that the
overall features of the locking range is still predicted well by
equation (6).
 
The above analysis suggests that whenever the interstitial 
vortex lattice is rectangular and the interstitial vortex interactions   
therefore cancel, Shapiro steps should be observed and be well
approximated by equation (6) when ${\rm sech}(\pi L_y/L_x)$ is small.
Square interstitial vortex arrangements are found at 
$B/B_{\phi} = 2, 1.5, 1.25, 1.0625$. For other filling fractions the
interstitial vortex lattice is not symmetrical and 
the interstitial-interstitial interactions do not cancel, leading to some
deviations from the predicted phase-locking (the locking range is
usually smaller than predicted). This is illustrated  
in Fig.\ 3 where we show the widths of the Shapiro steps for different 
filling fractions for a fixed AC amplitude and frequency. 
The Shapiro step widths for the different symmetrical vortex configurations at
$B/B_{\phi} = 2, 1.5$ and $1.0625$ are essentially identical.
For $B/B_{\phi} = 1.375$, and $1.68$ the steps are considerably reduced
and some 
{\it fractional} Shapiro steps also appear. We 
find in general that for $B_{\phi} < B < 2B_{\phi}$, the filling
fractions that produce square interstitial vortex lattices 
have the same Shapiro step widths as at $B = 2B_{\phi}$.   

Interestingly for $B > 2B_{\phi}$ we find that the step widths remain the
same as at $B = 2B_{\phi}$; 
however, there is a component in $V_{x}$ of the steps that  
linearly increases  with increasing $f_{d}$. For increasing magnetic  
fields this linear increase in $V_{x}$ of the steps increases until   
$B \geq 2.25B_{\phi}$, when the steps can no longer be
discerned. This linear increase suggests that only a portion of the
vortices are phase locked.
The images (not shown) from the simulations suggest that the extra 
vortices which have been added 
to the $B= 2B_{\phi}$ sub-lattice cause an additional soliton-like
motion which moves at a different speed than the 
interstitial vortices. To examine this we plot in figure 4 the time 
dependent vortex velocities for two separate interstitial vortices
along the same row at the $n = 1$ step ($f_{d} = 0.39$). 
In Figs.\ 4a and 4b for $B = 2B_{\phi}$ the signals for the two particles are
identical indicating that the vortices are moving in phase. 
In Figs.\ 4c and 4d we plot the signal from a row containing an extra vortex
for $B = 2.0625B_{\phi}$. Here the same oscillation
as in Figs.\ 4a and 4b is seen, indicating 
that phase-locking is occurring; however, there is an additional lower 
frequency oscillation superimposed. The soliton like nature of this 
disturbance  
can be seen by noting this extra oscillation out of phase between the
two vortices;   similar to a kink soliton on a
Frenkel-Kontorova chain \cite{FK}.

In conclusion, we have observed Shapiro steps in the current-voltage
characteristics of driven 
vortex lattices interacting with periodic pinning. 
At $B = 2B_{\phi}$ where the vortex motion 
consists of the one dimensional flow of interstitial vortices
between the pinned vortices, Shapiro 
steps are observed 
in agreement with recent experiments \cite{Bael}. 
We show that for certain filling fractions    
the equation of motion for a driven interstitial vortex
with a drive can be mapped to a driven overdamped pendulum. 
We derive the widths of the Shapiro steps as a function of relevant
experimental parameters, and find excellent agreement 
between theory and simulations. 
For filling fractions where interstitial-interstitial vortex interactions
become relevant the step widths are reduced. For $B > 2B_{\phi}$ the 
steps begin to vanish due to an additional soliton like flow
and other dynamical complexity.

{\bf Acknowledgments:}
We thank C.J.~Olson for critical reading of this manuscript.
This work was supported by the Director, Office of Advanced Scientific
Computing Research, Division of Mathematical, Information, and 
Computational Sciences of the U.S.\ Department of Energy under contract
number DE-AC03-76SF00098 as well as CLC and CULAR
(Los Alamos National Laboratory).

\begin{figure}
\center{
\epsfxsize=3.5in
\epsfbox{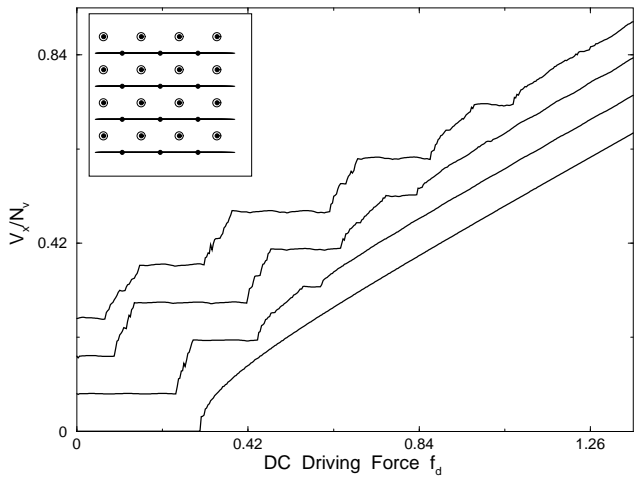}}
\caption{
Average vortex velocity $V_{x}$ versus applied DC driving for varying 
applied AC driving of (from top to bottom)
$ f_a = 0.814$, $0.5078$, $0.22135$, and $0$ for $B = 2B_{\phi}$. 
The curves have been shifted up from the $0$ curve for presentation.
Here clear steps can be seen for finite applied AC drives. Inset:
the typical vortex trajectories, positions (black circles) and pinning
sites (open circles) from the simulations. Here the motion consists of
interstitial vortices moving in one dimensional channels between the
vortices pinned at the pinning sites.  Parameters are: $\omega=1.6276$ and
$L_x=L_y=1.83$.
}
\label{fig2}
\end{figure}   

\begin{figure}
\center{
\epsfxsize=3.4in
\epsfbox{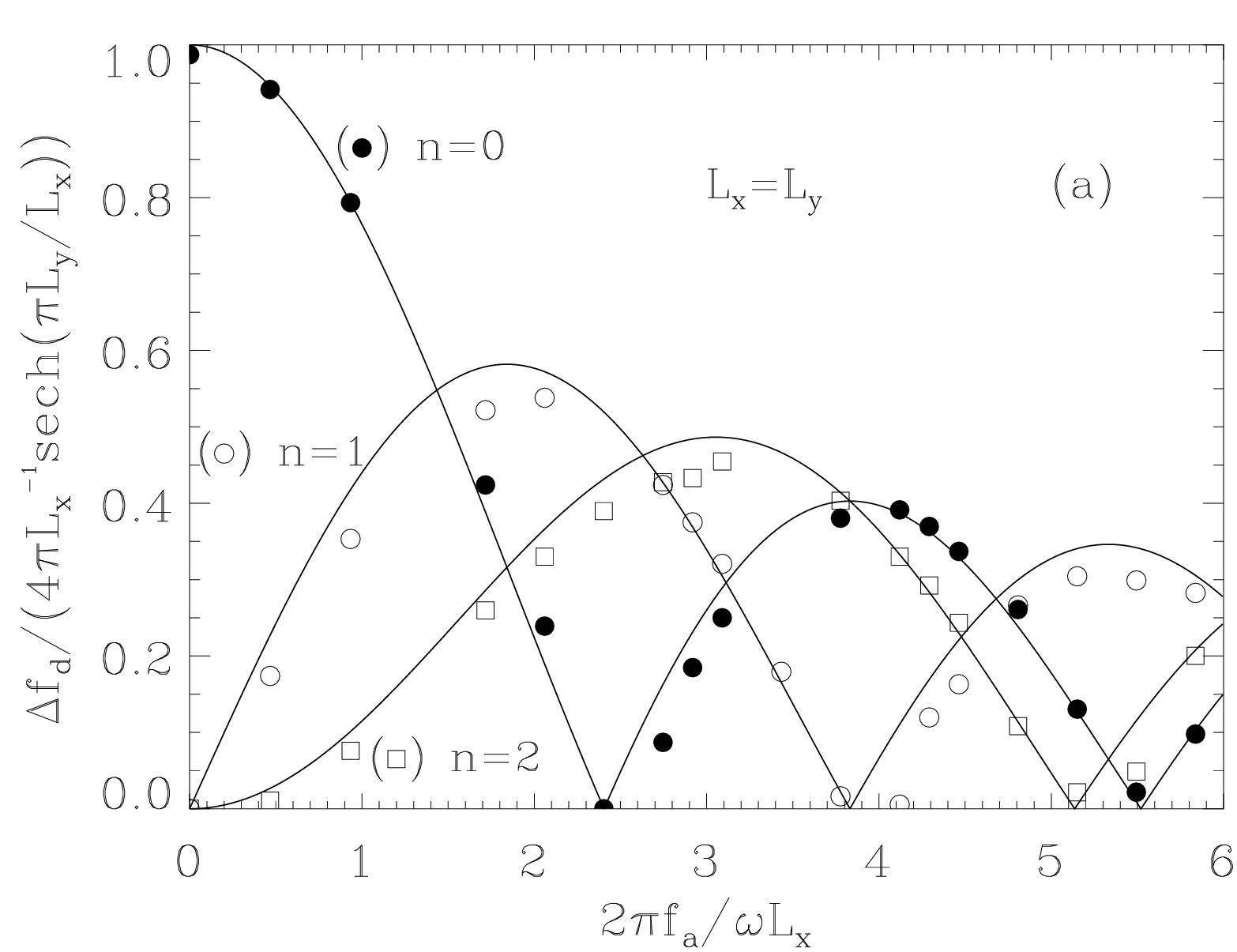}}
\center{
\epsfxsize=3.4in
\epsfbox{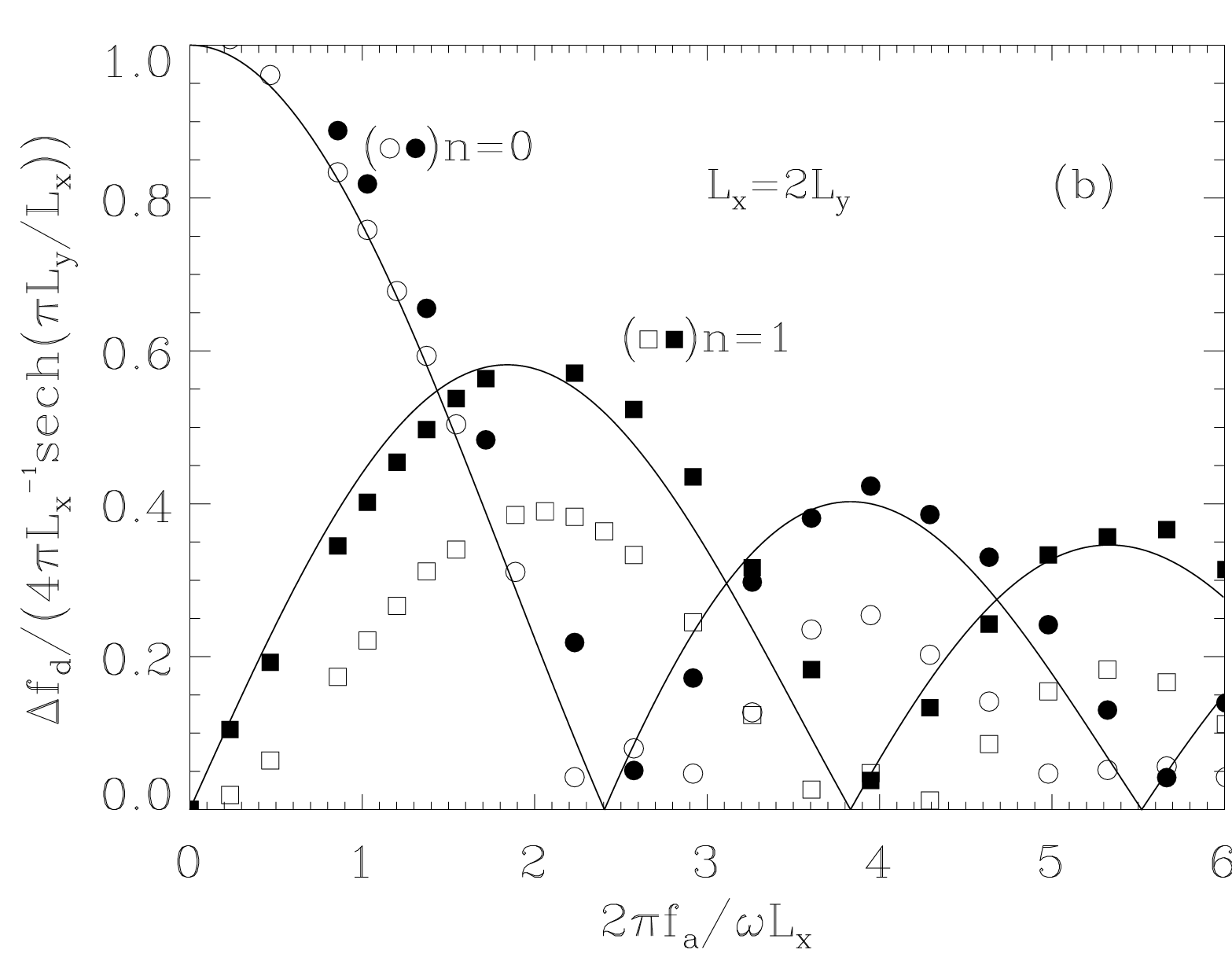}}
\caption{
The solid lines show the 
predicted dependence of the widths of the steps in the
velocity-force curves for varying applied AC drive 
amplitude from equation (6) for
the first three harmonics $n = 0$, $n = 1$, and $n = 2$.
Markers represent the simulation results of the phase-locking
range in DC current. $\omega=1.6276$.
AC amplitude, $f_a$, is varied from $f_a=0$ to
$f_a\approx 2.93\frac{L_x}{L_y}$.
(a) $L_x = L_y = 1.83$ and $B=2B_{\phi}$.
(b) $L_x = 2L_y = 3.66$; open markers represent
simulations at $B=2B_{\phi}$ and filled makers represent simulations
with a single unpinned vortex.}
\label{fig3}
\end{figure}

\begin{figure}
\center{
\epsfxsize=3.5in
\epsfbox{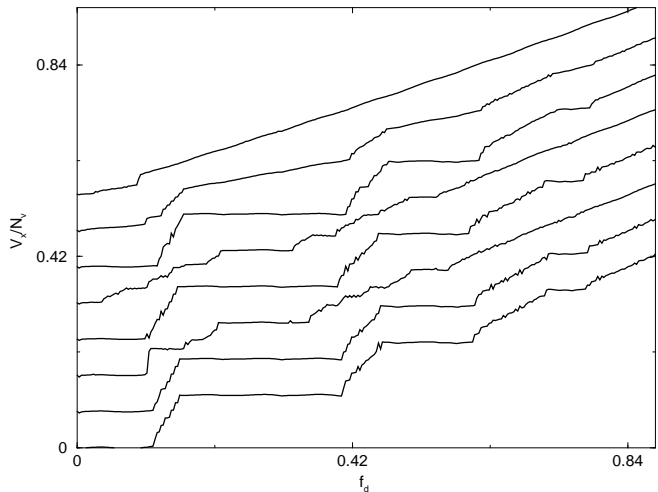}}
\caption{ 
Average vortex velocity $V_{x}$ versus driving force for constant AC driving   
$f_a = 0.5078$ for varying fields (from top to bottom) 
$B/B_{\phi} = 2.25, 2.125, 2.0168, 1.5, 1.375, 1.25, 1.0625$ .
Each curve has been shifted for clarity. Parameters are: $\omega=1.6276$ and
$L_x=L_y=1.83$.}
\label{fig4}
\end{figure}  


\begin{figure}
\center{
\epsfxsize=3.5in
\epsfbox{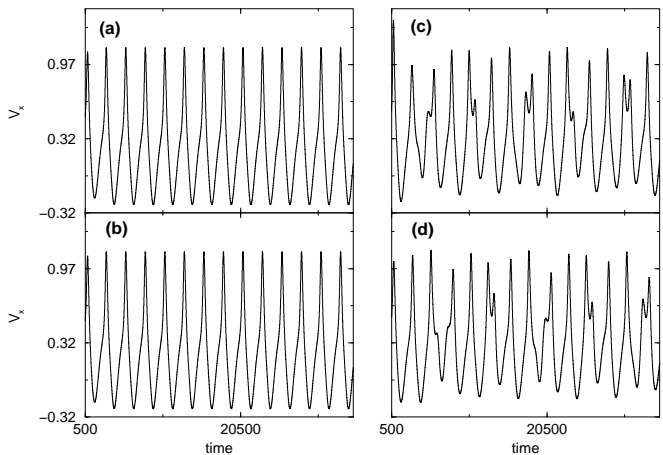}}
\caption{
(a) The time dependent vortex velocity for two different vortices
in a single row for  $B = 2B_{\phi}$ in the middle of the $n=1$ step for
the same system as in Fig.\ 1. (b) The same for $B = 2.0625B_{\phi}$.
Parameters are: $\omega=1.6276$, $L_x=L_y=1.83$, $f_d=0.38$, and 
$f_a=0.442$.}
\label{fig4a}
\end{figure}  

\end{document}